\documentclass[nocompress]{spie}  
\usepackage[]{xspace,amsmath,graphicx,hyperref}
\usepackage[caption=false]{subfig}
\newcommand{\FIG}[3]{\includegraphics[width=#1\linewidth,draft=#2]{#3.eps}}
\newcommand{\cre}[1]{{\small\textit{---~Credit: #1}}}
\newcommand{\mum}{\mbox{{\usefont{U}{eur}{m}{n}{\char22}}m}\xspace}

\graphicspath{{D:/Documents/5_CASSINI/FIGURES/}}

\title{Could Jean-Dominique Cassini see the famous division in Saturn's rings?} 

\author{Julien Lozi\supit{a,b,c}, Jean-Michel Reess\supit{d}, Alain Semery\supit{d}, Emilie Lhom\'e\supit{d}, Sophie Jacquinod\supit{d}, Michel Combes\supit{d}, Pernelle Bernardi\supit{d}, R\'emi Andretta\supit{d}, Maxime Motisi\supit{d}, Laurence Bobis\supit{e} and Emilie Kaftan\supit{e}
\skiplinehalf
\supit{a}University of Arizona, 1401 E University Blvd, Tucson, AZ 85721, USA; \\
\supit{b}CNES, 18 avenue Edouard Belin, 31400 Toulouse, France; \\
\supit{c}Onera, BP72 -- 29 avenue de la Division Leclerc, 92322 Ch\^atillon Cedex, France; \\
\supit{d}LESIA -- Observatoire de Paris, 5 place Jules Janssen, 92195 Meudon Cedex, France; \\
\supit{e}Observatoire de Paris, 61 avenue de l'Observatoire, 75014 Paris, France
}

\authorinfo{Further author information: (Send correspondence to J.-M.R.)\\ J.L.: E-mail: jlozi@email.arizona.edu, Telephone: 1 650 604 6549 \\J.-M.R.: E-mail: jean-michel.reess@obspm.fr, Telephone: +33 (0)1 45 07 76 58}

\begin{document} 
  \maketitle 

\begin{abstract}
\end{abstract}

Nowadays, astronomers want to observe gaps in exozodiacal disks to confirm the presence of exoplanets, or even make actual images of these companions. Four hundred and fifty years ago, Jean-Dominique Cassini did a similar study on a closer object: Saturn. After joining the newly created Observatoire de Paris in 1671, he discovered 4 of Saturn's satellites (Iapetus, Rhea, Tethys and Dione), and also the gap in its rings. He made these discoveries observing through the best optics at the time, made in Italy by famous opticians like Giuseppe Campani or Eustachio Divini. But was he really able to observe this black line in Saturn's rings? That is what a team of optical scientists from Observatoire de Paris - LESIA with the help of Onera and Institut d'Optique tried to find out, analyzing the lenses used by Cassini, and still preserved in the collection of the observatory. The main difficulty was that even if the lenses have diameters between 84 and 239~mm, the focal lengths are between 6 and 50~m, more than the focal lengths of the primary mirrors of future ELTs. The analysis shows that the lenses have an exceptionally good quality, with a wavefront error of approximately 50~nm rms and 200~nm peak-to-valley, leading to Strehl ratios higher than 0.8. Taking into account the chromaticity of the glass, the wavefront quality and atmospheric turbulence, reconstructions of his observations tend to show that he was actually able to see the division named after him.



\section{INTRODUCTION}
\label{sec:intro}

The observation of exoplanets today gives the same level of discovery as the observation of the solar system with the first astronomical instruments. We can compare the discovery of exoplanets to the discovery of natural satellites in the solar system, and the observation of Saturn's ring to the observation of protoplanetary disks and exozodiacal light.

Towards the end of the \textsc{xvi}\textsuperscript{th}~century, were invented the first magnifying lenses, for military observation applications. It was not until the year 1609 that these lenses are turned toward the sky: Galileo is one of the precursors, especially when he presented his refractive telescope in August 21, 1609, in front of the Doge and the members of the Venetian Senate at the campanile of St Mark's Square. His first objective lenses had diameters between 30 and 70~mm, but had a poor quality, both in refractive index uniformity and surface roughness. According to him, among his 60~lenses, only a few were good enough to be used.

Despite the poor quality of his instruments, in July 25, 1610, Galileo observed Saturn's rings, but he was not able to understand their geometry. Huygens was the first to understand their true nature in 1656, describing: "It is surrounded by a thin, flat ring which nowhere touches the ground and is inclined to the ecliptic\cite{Huygens1659}."

The quality of the telescopes improved significantly before the end of the \textsc{xvii}\textsuperscript{th}~century, especially with the improved techniques used by French and Italian opticians\cite{Miniati02}. Jean-Dominique Cassini enjoyed those quality optics, and obtained a few discoveries, especially the division in Saturn's rings. But  was he really able to observe the division with his telescopes? We tried to answer this question by analyzing a few of his objective lenses, and simulate the quality of his observations.

\section{A brief history}
\label{sec:ABriefHistory}

\begin{figure} \centering
  \FIG{0.4}{false}{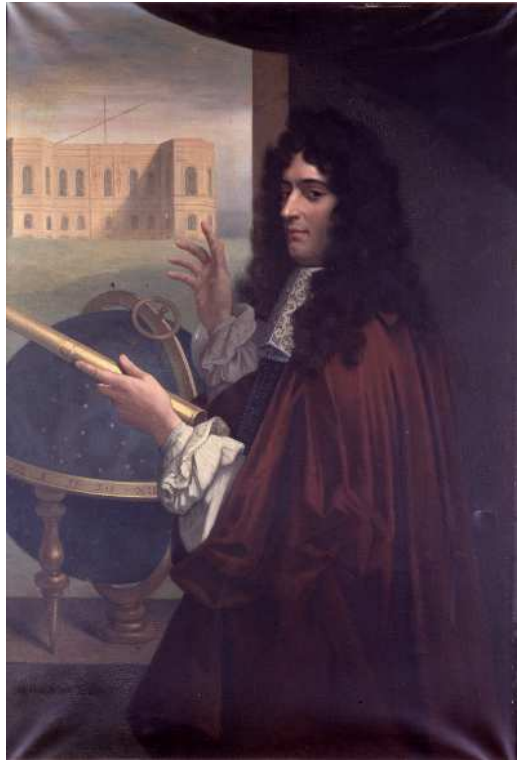}
  \caption{Portrait of Jean-Dominique Cassini by L\'eopold Durangel (1879), with the facade of Paris Observatory in the background, together with the 34-foot telescope on its roof \cre{Library of Paris Observatory}.}
  \label{fig:cassini}
\end{figure}

In 1669, Jean-Dominique Cassini is called in France by the Minister of Finances Jean-Baptiste Colbert to participate in the newly created Academy of Sciences. In 1671, with a grant of Louis XIV, he participated in the creation of the Paris Observatory. Cassini brought from Italy a 17-foot\footnote{The size of the foot used here is unclear. It could be the Italian "pied de Rome" (297.896~mm), or the French "pied du Roi" (324.84~mm), or even another unit of length, as there was dozens of definitions at that time. However, as we will see in Sec.~\ref{sec:MechanicalCharacteristics}, it is more likely that it corresponds to the "pied du Roi". In this case, 17~feet $\approx$ 5.5~m.} refractive telescope made ​​by Campani, famous optician and astronomer from Rome. With this telescope, manufactured before 1665, Cassini has discovered the eclipses of the Galilean satellites, and the Great Red Spot. He calculated the rotational period of Jupiter in 1665 and Mars in 1666\cite{Cassini1705}. At the creation of Paris Observatory, Colbert asked Campani to provide his best lens, and to make others with larger focal lengths. In December 1672, Campani sent a 34-foot lens (11~m), and later lenses of 80, 90, 100 and 136 feet (26, 29, 32 and 44~m)\cite{Wolf1902}. The 34-foot lens was mounted on a 11-meter tube, and moved by a pulley system to point objects. This telescope is visible on the top of Paris Observatory in the background of Cassini's portrait (Figure~\ref{fig:cassini}).

With this telescope, he described in 1675 a separation in Saturn's rings, which will be later named the Cassini Division\cite{Cassini1676,Cassini1677}:

\begin{quotation}
Deinde latitudo Annuli dividebatur bifariam, Line\^a obscur\^a apparenter Elliptic\^a rever\^a Circulari quasi in duos annulos concentricos, quorum interior exteriori lucidior erat. Hanc phasim statim post emersionem Saturni \`e Solaribus radiis per totum annum usque ad ejus Immersionem conspexi ; prim\`o quidem, Telescopie Pedum 35, deinde minori, Pedum 20. Ejus delineationem, utcumque rudem, properante calamo hic adjeci.

Then the width of the ring was divided by a dark line, which looks elliptical, but actually circular, as two concentric rings, the inner ring being brighter than the outer ring. I observed throughout the year immediately after this phase where Saturn emerged from the Sun's rays until its immersion; first with the 35-foot telescope, then with smaller 20-foot one. I sketched it, although roughly, and have added hastily here.
\end{quotation}

\emph{An Extract of Signor Cassini's Letter concerning a Spot lately seen in the Sun, together with a remarkable Observation of Saturn, made by the same}, Philosophical Transactions, \textbf{Vol.~XI}, September 25\textsuperscript{th}, 1676, p.~690.

At the time of his observations, the rings had a diameter of 38.5~arcsec, and the Cassini division had a maximum width of 0.65 arcsec. With a pupil diameter of 108~mm, the theoretical resolution of the 34-foot telescope is approximately 1~arcsec. Combined with the lens chromaticity, its defects and the atmospheric turbulence, there was always a doubt that Cassini has actually observed the division.

\begin{figure} \centering
  \FIG{0.9}{false}{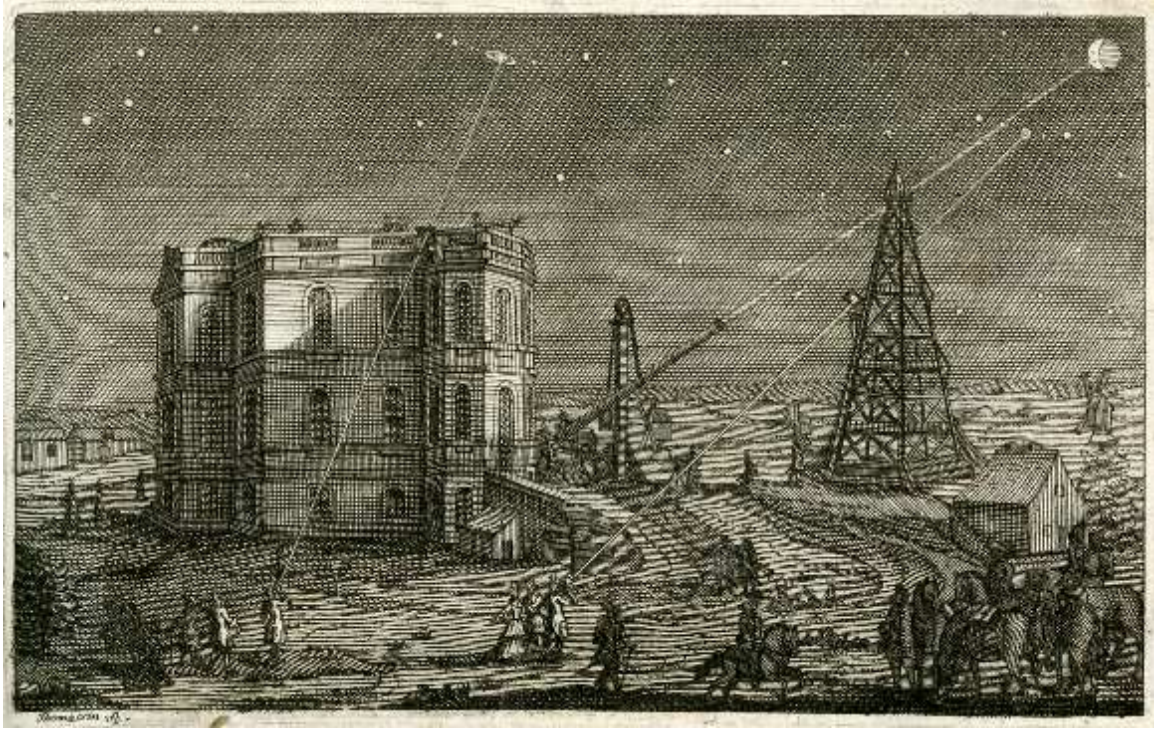}
  \caption{Observing night at Paris observatory, around 1691 \cre{Library of Paris Observatory}.}
  \label{fig:obs-paris}
\end{figure}

The other lenses have very large focal lengths, up to 50~m, which is why they could not use tubes and pulleys\cite{Prince1882}. Figure~\ref{fig:obs-paris} is an engraving of an observation at Paris Observatory, seen from the west side of building around 1691\cite{Cassini1691}. On the left, we can see an astronomer observing Saturn surounded by 4~satellites, with a lens placed within 100~feet on the roof of the Observatory; on the center, the 34-foot telescope is used to look at the Moon; on the right, an observation of Jupiter and its satellites is made with a long focal lens placed on the Marly tower, a military tower repatriated at the request of Marquis de Louvois, minister of Louis XIV.

These lenses are still stored at Paris Observatory. Following a request from the Observatory curators, a group of optical engineers from LESIA (Meudon) and Onera (Ch\^atillon) was created to study these lenses. In a first study, we tested the five largest lenses, including the famous 34-foot lens. The results of this study were presented in 2012 during the exhibition dedicated to the tercentenary of Cassini's death\footnote{Website (in French): \url{http://expositions.obspm.fr/cassini/index.php}\\A video presenting our experiment can be found here (in French): \url{http://www.youtube.com/watch?v=0LWtQvvdb4I}}.

A similar study was made a few years ago concerning objective lenses, eyepieces and complete telescope made by Torricelli, Divini and Campani\cite{Vanhelden99,Baiada95}. Results were presented in\cite{Miniati02,Bonoli02}. Also, simulations of Cassini have already been made using a few hypotheses about the quality of the 34-foot lens, leading to a conclusion different from ours\footnote{{Oncle Dom}, "L'affaire Cassini", Sept. 2005\\
\url{http://oncle.dom.pagesperso-orange.fr/astronomie/histoire/anneaux/structure/cassini/cassini.htm}}.


\section{Analysis of the lenses}
\label{sec:AnalysisOfTheLenses}

\subsection{Mechanical characteristics}
\label{sec:MechanicalCharacteristics}

\begin{figure} \centering
  \subfloat[Lenses used by Cassini \cre{J. Counil / Library of Paris Observatory}.]{\label{fig:objectifs}
    \FIG{0.8}{false}{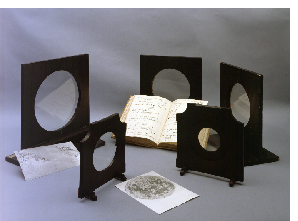}}\\
  \subfloat[Engraving of optician Campani on the 34-foot lens.]{
    \label{fig:C40} \FIG{0.8}{false}{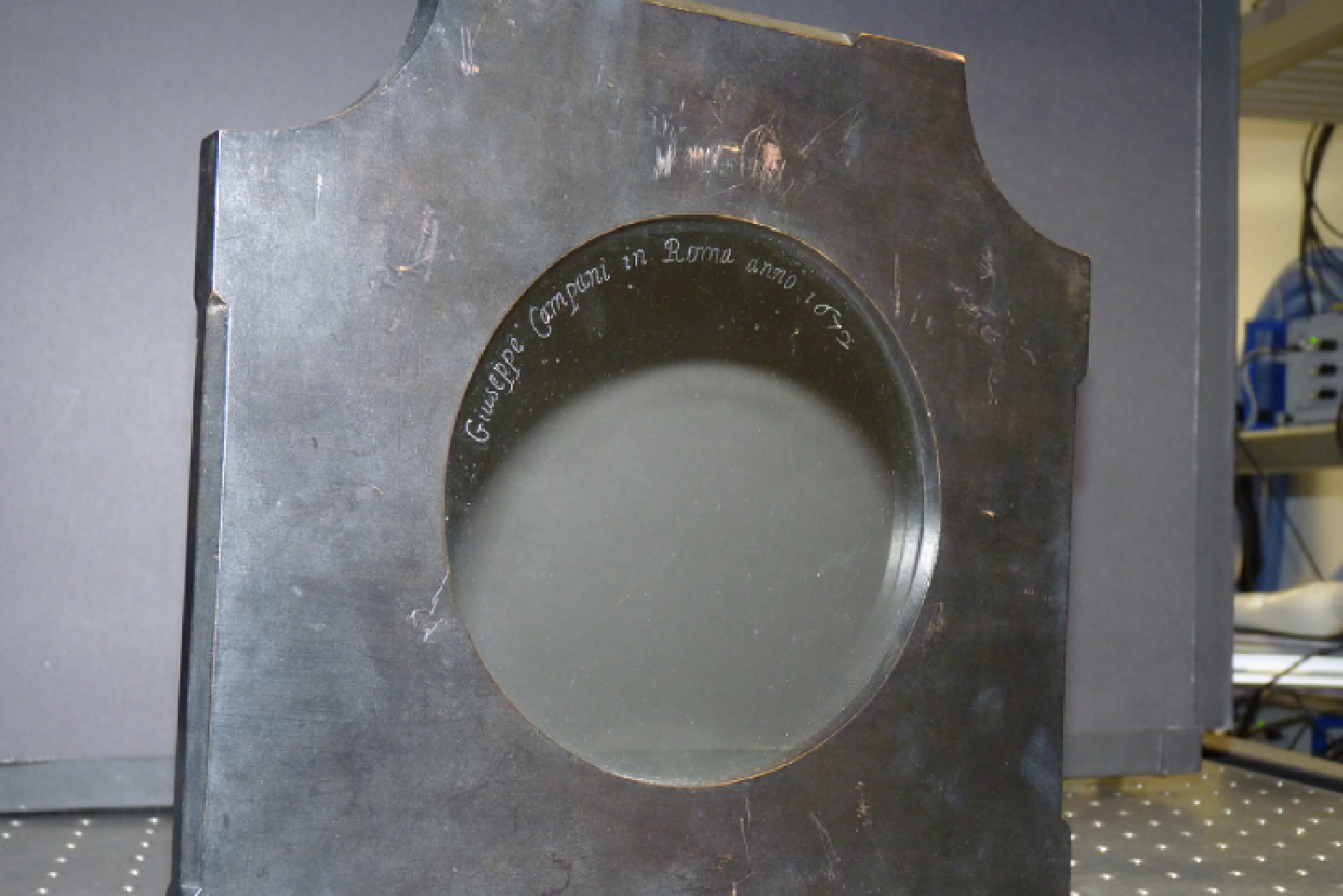}}
  \caption[Cassini lenses.]{Cassini lenses, with at least two made by Campani in Rome.}
  \label{fig:obj-C40}
\end{figure}

Paris Observatory has a collection of more than thirty lenses from Cassini's period. However, we only know very little about them, except a few rough focal lengths. But the precise focal lengths, the glass quality and chromatic variations in the index are unknown. These data are essential to determine the quality of Cassini's observations, and more precisely to know if he observed the division in Saturn's rings. This is why we have studied the characteristics of the five largest, mounted in wooden frames, and shown in Fig.~\ref{fig:obj-C40}.

The five studied lenses are shown in Fig.~\ref{fig:objectifs}, and the engraving on the 34-foot lens is detailed in Fig.~\ref{fig:C40}: \emph{Giuseppe Campani in Roma anno 1672\/}. In the rest of this article, the five lenses will be named after their referencing number. On Fig.~\ref{fig:C40}, bottom left to right: lenses \#40 (the 34-foot lens) and \#44, in the top left right: lenses \#42, \#43 and \#41. They all have diameters greater than 135~mm, but some are vignetted either by the wooden frame or by a cardboard mask, certainly used to limit aberrations. Only lenses \#40 and \#41 have a visible engraving by Campani.

First, we want to determine the mechanical properties of the lenses, i.e. their diameters, thicknesses and the faces radii of curvature. The pupil diameters $D$ and thicknesses $e$ are measured using a caliper and a positioning sensor. The curvatures are calculated using a spherometer. It consists of a circle of radius $r$, and a central positioning sensor which measure the difference $h$ between the surface and a reference plane. The radius of curvature is then given by the equation:
\begin{equation}
  R = \frac{r^2+h^2}{2h}.
\end{equation}

Table 1.2 summarizes the characteristics of the five lenses. Focal lengths have been measured with a laser at 532~nm. Chromaticity is very important, so these values ​​vary greatly with the wavelength, over several centimeters. There are for example 50~mm between the focal planes at 633 nm and at 532~nm for the lens \#40.

\begin{table} \centering
  \caption{Table summarizing the characteristics of the lenses, for their two faces.}
  \medskip
  \begin{tabular}{rccccc}
    \hline\hline
    Inventory number & 40 & 41 & 42 & 43 & 44 \\
    \hline
    $D$ [mm] & 137 & 181 & 239 & 183 & 84 \\
    \hline
    $r$ [mm] & 45 & 75 & 75 & 75 & 35 \\
    $h_1$ [\mum{}] & $85\pm5$ & $75\pm5$ & $22\pm5$ & $-15\pm5$ & $-64\pm5$ \\
    $h_2$ [\mum{}] & $85\pm5$ & $61\pm5$ & $105\pm5$ & $120\pm5$ & $245\pm5$ \\
    \hline
    $R_1$ [m] & $11.9\pm0.7$ & $37.5\pm2.5$ & $128\pm30$ & $-188\pm70$ & $-9.57\pm0.8$ \\
    $R_2$ [m] & $11.9\pm0.7$ & $46.1\pm3.8$ & $26.8\pm1.3$ & $23.4\pm1.0$ & $2.50\pm0.05$ \\
    $e$ [mm] & $6.25$ & --- & --- & --- & --- \\
    \hline
    $f_\text{cal @ 532~nm}$ [m] & $11.2\pm0.7$ & $39.0\pm2.9$ & $41.7\pm3.3$ & $50.5\pm4.9$ & $6.39\pm0.35$ \\
    $f_\text{cal @ 532~nm}$ [feet] & $34.6\pm2.0$ & $120\pm9$ & $129\pm10$ & $155\pm15$ & $19.7\pm1.1$ \\
    \hline
    $f_\text{mes @ 532~nm}$ [m] & $10.9$ & $40.2$ & $47.3$ & $48.5$ & $6.30$ \\
    $f_\text{mes @ 532~nm}$ [feet] & $33.6$ & $123.8$ & $145.6$ & $149.3$ & $19.4$ \\
		\hline
		f-number & 80 & 222 & 198 & 265 & 75 \\
    \hline\hline
  \end{tabular}
  \label{tab:car-lentilles}
\end{table}

The thicknesses $e$ have not all been measured, because we cannot remove the lenses from their wood frames for preservation. However, we can say that all lenses are very thin, with a ratio $D/e$ around 20, whereas the modern lens have this ratio smaller than 10, to avoid deformations. In the days of Cassini, the inhomogeneities of the glass were important, thus minimizing the thickness reduced their impact.

In Tab.~\ref{tab:car-lentilles}, we can see that lenses \#42 and \#43 are almost plano-convex, while the lens \#40 is symmetric. Lens \#44 is a positive meniscus. The calculated focal lengths $f_\text{ca@ 532~nm}$ were obtained using a typical crown glass index of $n=1.53$. They match approximately the measured focal length $f_\text{me@ 532~nm}$. Also, the focal length of the 34-foot lens is 10.9~m, so it validates the hypothesis that the foot used by Campani and Cassini was the French "Pied du Roi", corresponding to 324.84~mm. The other focal lengths does not match the lenses described in~\ref{sec:ABriefHistory}, especially lens \#44 is clearly not the 17-foot lens brought back form Italy by Cassini, but rather probably the 20-foot lens mentioned by Cassini in the citation of Sec.~\ref{sec:ABriefHistory}.

\subsection{Wavefront measurement}
\label{sec:WavefrontMeasurement}

These characteristics we obtained previously are not enough to conclude about the quality of Cassini's observations. To characterize an optical system, it is important to measure the wavefront. We used a Zygo interferometer, with a spherical mirror at the focal point of the lens. But for these lenses, with some focal length greater than 40~m, it is very difficult to put the spherical mirror at the focal point, and over such distances, atmospheric turbulence deteriorates a lot the measurement. So we used a reference lens with a focal length of 750~mm, placed just after the lens we want to measure. This doublet, whose characteristics are known, reduced the focal length of all the lenses to less than a meter, making the measurement more convenient.

However, the doublet was designed and optimized for a collimated beam, and here we are using it in a convergent beam after the historical lenses. So it brings a spherical aberration, which was corrected from the measurements in post treatment. Some of the lenses have diameters greater than the pupil of the Zygo, so we measured the wavefront of different zones of the lenses, and stitched them together afterwards.

\begin{figure}[b] \centering
  \FIG{0.7}{false}{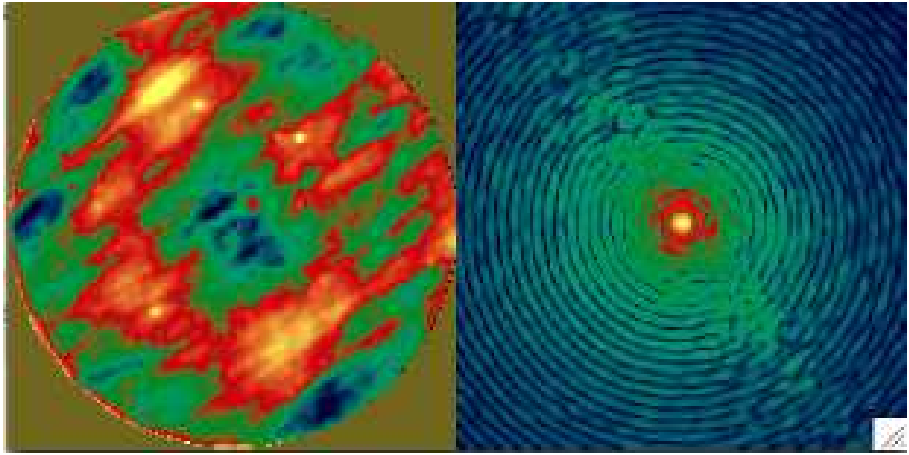}
  \caption{Wavefront measurement and calculated PSF of the 34-foot lens.}
  \label{fig:res-40}
\end{figure}

With the wavefront, we deduced the point spread function (PSF) of each lens. The results of the 34-foot lens (\#40), the one used to observe the division in Saturn's rings, are presented in Fig.~\ref{fig:res-40}.

\begin{figure} \centering
  \subfloat[Wavefront and PSF for the lens \#41.]{\label{fig:res-41}
    \FIG{0.7}{false}{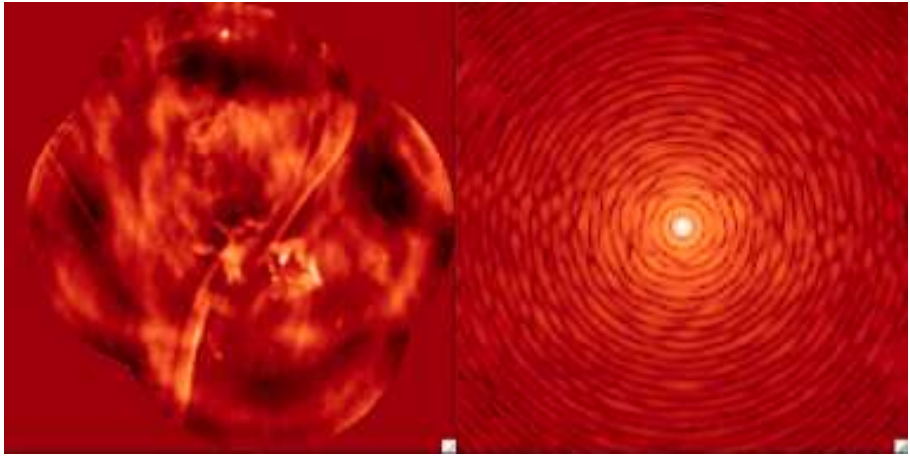}}\\
  \subfloat[Wavefront and PSF for the lens \#43.]{\label{fig:res-43}
    \FIG{0.7}{false}{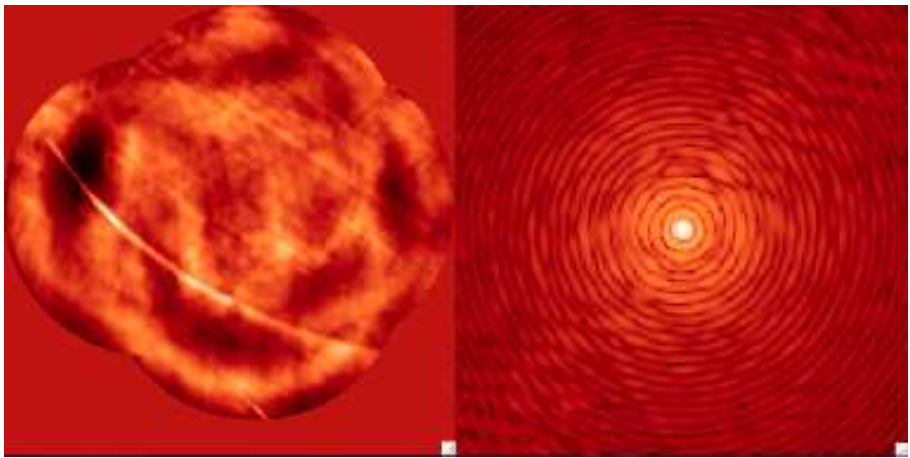}}\\
  \subfloat[Wavefront and PSF for the lens \#44.]{\label{fig:res-44}
    \FIG{0.7}{false}{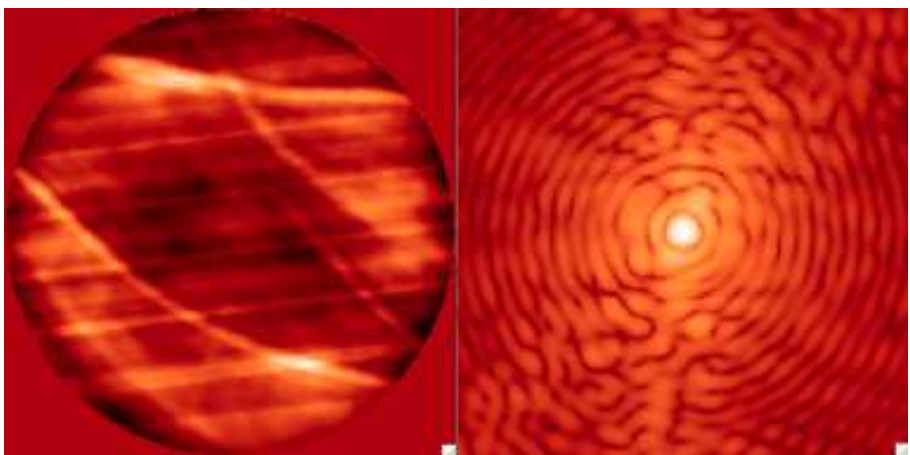}}\\
  \caption{Wavefront measurements and calculated PSF for the lenses \#41, \#43 et \#44.}
  \label{fig:res-obj}
\end{figure}

With a amplitude peak-to-valley of $0.325\;\lambda$ @ $\lambda = 632.8$~nm and a standard deviation of $\boldsymbol{0.049\;\lambda}$, the optical quality is very good for a \textsc{xvii}\textsuperscript{th}~century instrument. With this wavefront error, the Strehl ratio is $\boldsymbol{0.83}$, so this lens provides images with a very good quality. We can see however a pattern in the wavefront error, with diagonal lines across the pupil. These are certainly index variations in the lens, created by the imperfect homogenization of the glass during its manufacture. Indeed, this pattern is typically due to the molten glass being stretched over and over to obtain an homogeneous index.

Figure~\ref{fig:res-obj} shows the wavefront measurements and calculated PSFs for three of the four remaining lenses: \#41, \#43 and \#44. Lenses \#41 to \#43 have diameter greater than the pupil of the Zygo, so we stitched different measurements using the local defaults in the wavefront. This reconstruction was not possible for the lens \#42, so it is not presented in Fig.~\ref{fig:res-obj}. The reconstruction does not cover the entire lens, which gives non-circular pupils for figures~\ref{fig:res-41} and~\ref{fig:res-43}.

We can also see that the correction of the spherical aberration of the reference lens is not perfect, because there is still a residue visible on the reconstruction, that is not due to the measured lens.

For the lenses \#41 and \#43, the wavefront quality is exceptional, with the same standard deviation of $\boldsymbol{0.027\;\lambda}$, which gives a Strehl ratio of $\boldsymbol{0.94}$. As for the 34-foot lens, variations of the wavefront are due to inhomogeneities of the glass left during its manufacture.

The lens \#44 is a bit worse than the previous ones: the wavefront has a standard deviation of $\boldsymbol{0.070\;\lambda}$, which gives a Strehl ratio of $\boldsymbol{0.67}$. The quality of the glass is also limited by inhomogeneities. This lens has its diameter highly reduced by a cardboard mask, which suggests that the defects are more important at its edges.

\subsection{Chromaticity of the refractive index}
\label{sec:ChromaticityOfTheRefractiveIndex}

The wavefront measurements were obtained with a monochromatic light at 633~nm. However, before the invention of the achromatic doublet in~1758, the main limitation of the lenses was the chromaticity. The only way to reduce it was to use long focal lengths. But even with a focal length of a dozen of meters, the chromaticity is still a major limiting factor in the image quality.

To measure the refractive index of the lenses, we used a white source and selected the wavelength with a monochromator. The source is placed at a long distance from the studied lens. Once again we used also the 750~mm reference lens to reduce the total focal length. Then we measured the position of the focal plane at different wavelength, using a microscope lens and an eyepiece. The measurement is made manually and visually, so the results are not very accurate, and depend on the observer. The measurements are then entered in an optical simulator, to take into account the chromaticity of the reference lens. 

\begin{figure}[b] \centering
  \subfloat[Measurement of the refractive index of the 34-foot lens, compared to a few crown glasses.]{\label{fig:index-40}
    \FIG{0.5}{false}{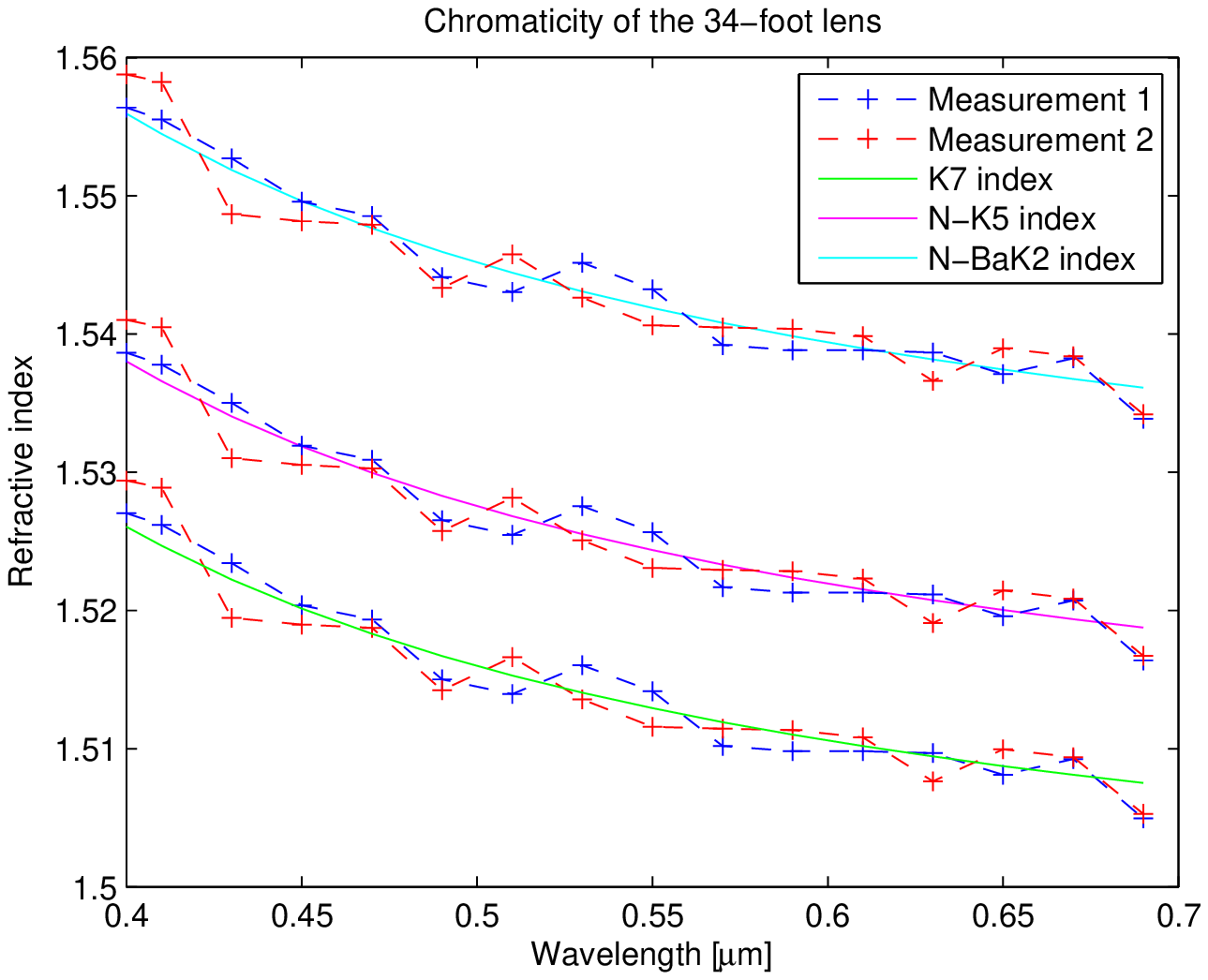}}
  \subfloat[Measurement of the refractive index of the lenses.]{\label{fig:index-lenses}
    \FIG{0.5}{false}{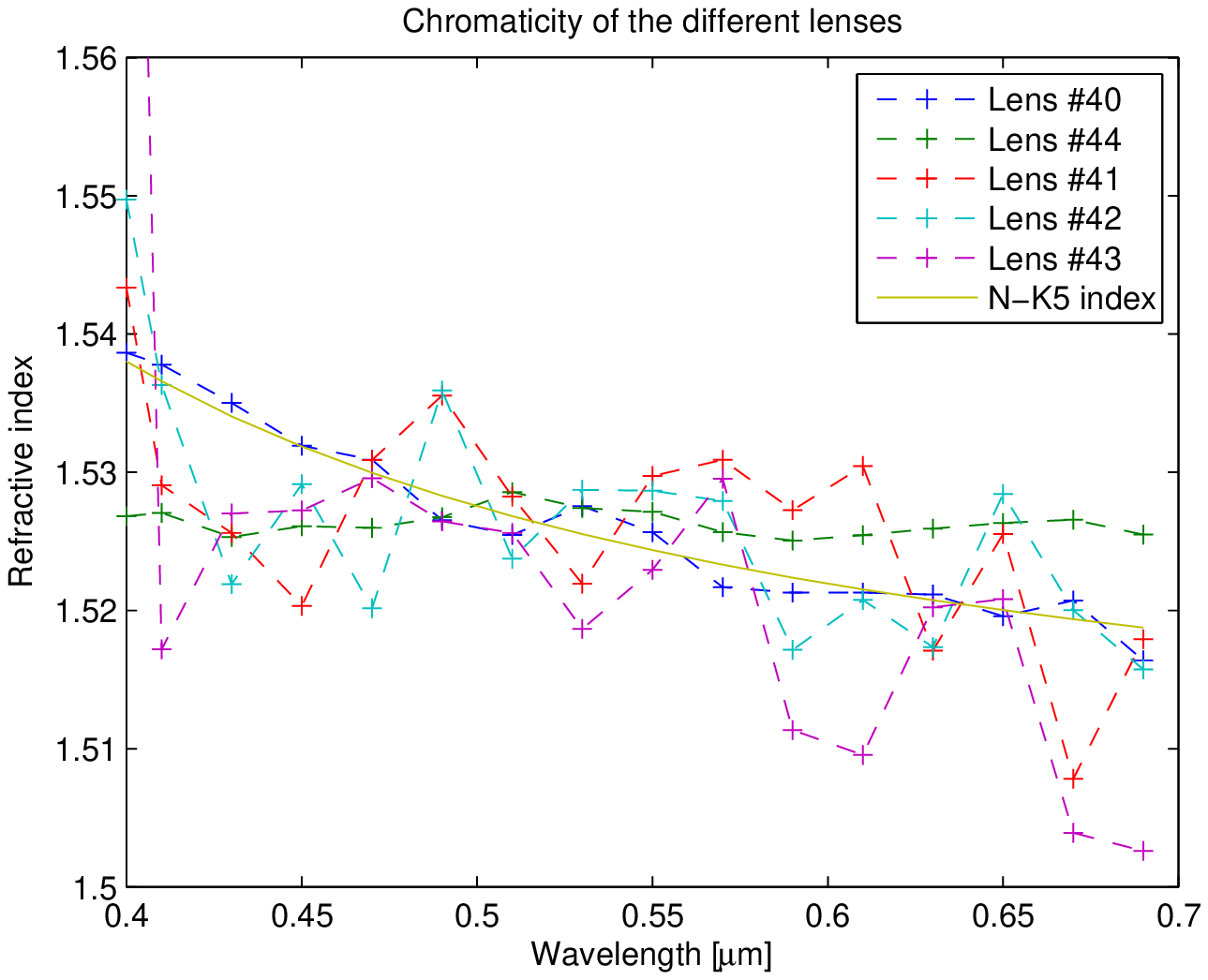}}
  \caption{Measurements of the refractive index.}
  \label{fig:index}
\end{figure}

Figure~\ref{fig:index} presents the measurements for the different lenses. On figure~\ref{fig:index-40}, we compare the two measurements made of the 34-foot lens with different modern crown glasses: Schott's K7, N-K5 (crowns) and N-BaK2 (barium crown). Due to the inaccuracy of the method, the mean refractive index is not well known, but with the measurement presented in Tab.~\ref{tab:car-lentilles}, it seems to be between 1.52 and 1.55, which are typical values for crown glasses. However on Fig.~\ref{fig:index-40}, we can see that the measured dispersion follows pretty well the typical dispersion of crown glasses, so the Abbe number of this lens  is around 60.

On Fig.~\ref{fig:index-lenses}, we compare the measurements of the different lenses to a crown N-K5. We can see that for the lenses~\#41, \#42 and \#43, i.e. the lenses with the highest focal lengths, the measurements are not precise enough to conclude on the index dispersion. Nevertheless, for lens~\#44, the refractive index is unexpectedly almost flat on the entire spectrum. This would mean that the lens is almost achromatic. There might be a problem in our measurements for this one, we might have to do this experiment again.


\section{What could Cassini see?}
\label{sec:WhatCouldCassiniSee}

\begin{figure} \centering
  \subfloat[Planets through the objective lens \#40.]{\label{fig:pla-40}
    \FIG{0.75}{false}{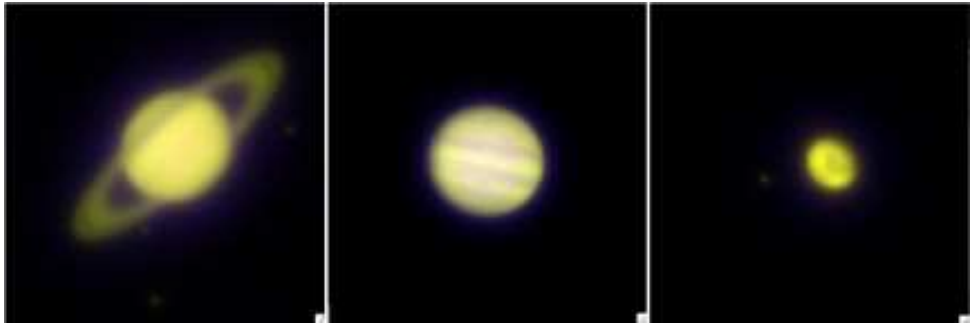}}\\
  \subfloat[Planets through the objective lens \#41.]{\label{fig:pla-41}
    \FIG{0.75}{false}{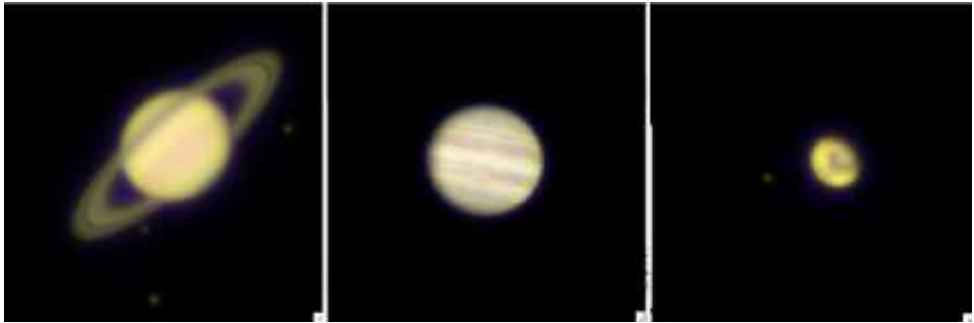}}\\
  \subfloat[Planets through the objective lens \#43.]{\label{fig:pla-43}
    \FIG{0.75}{false}{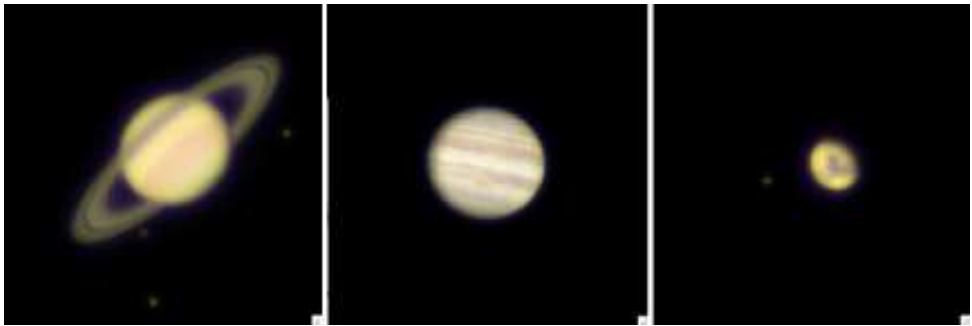}}\\
  \subfloat[Planets through the objective lens \#44.]{\label{fig:pla-44}
    \FIG{0.75}{false}{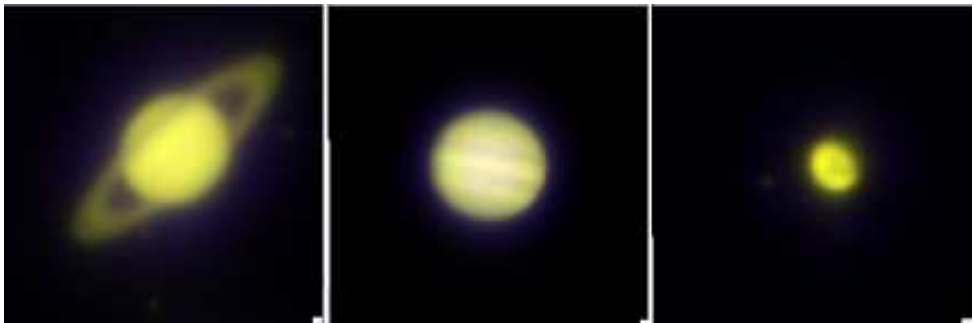}}\\
  \caption{Saturn, Jupiter and Mars seen through lenses \#40, \#41, \#43 and \#44.}
  \label{fig:pla-obj}
\end{figure}

The results presented in Sec.~\ref{sec:WavefrontMeasurement} apply only to the objective lenses, however almost nothing is known about the eyepieces Cassini used. They can also bring aberrations, but if they have the same optical quality than the objective lenses,  their impact should be limited. Therefore, for the simulations presented in Fig.~\ref{fig:pla-obj}, we assumed perfect eyepieces with the same refractive index as the objective lenses. Using a planetarium software, we simulated images of Saturn, Jupiter and Mars as they were at the beginning of 1675. Then, using the mechanical characteristics presented in section~\ref{sec:MechanicalCharacteristics}, the PSFs calculated in Sec.~\ref{sec:WavefrontMeasurement}, the chromaticity measured in Sec.~\ref{sec:ChromaticityOfTheRefractiveIndex}, and the spectral sensitivity of the human eye, we obtained the images presented in Fig.~\ref{fig:pla-obj}.

The satellites visible in these images do not have the right magnitude, and should not be taken into account. The chromatic aberration is clearly visible on these simulations, especially in the blue wavelengths. In fact, when the telescope is focused at 550~nm --- the maximum
sensitivity of the human eye --- the image in the blue is totaly out of focus, especially for
the lowest focal lengths of lenses \#40 and \#44. Even if it is very difficult with lens \#44,
we can definitely see the division in Saturn's rings, as well as the Great Red Spot on Jupiter and dark areas on Mars. 

These simulations do not take into account the atmospheric turbulence. We performed dynamical simulations, using bad seeing conditions that Paris should have in the \textsc{xvii}\textsuperscript{th}~century: we used a Fried parameter of 30 to 50~mm, and a wind of 5~km.s$^{-1}$. The results are presented here\footnote{Video for a Fried parameter of 30~mm: \url{http://www.youtube.com/watch?v=eZiIgcBwnSI}\\Video for a Fried parameter of 50~mm: \url{http://www.youtube.com/watch?v=rvm2xoIuAdQ}}. The Fried parameter is the same order of magnitude than the pupil dimensions, so the impact of the turbulence is more global than local, the image is sometimes mostly blurry, sometimes it is mostly clear. So we can still see the division.


\section{Conclusion}
\label{sec:Conclusion}

For resolved objects like the planets of the solar system, a classical definition of the angular resolution is not sufficiant to conclude on Cassini's discovery. Even with complex simulations like the one we presented in Sec.~\ref{sec:WhatCouldCassiniSee}, it is impossible to reproduce the sensation of a direct observation. The eye is a powerful instrument, because it is able to integrate information up to 6~seconds, keeping the most important details of pictures, similar to the modern technique of lucky imaging\cite{Fried78}. With this property and the simulations presented in Sec.~\ref{sec:WhatCouldCassiniSee}, it is most likely that Cassini had observed the division with one of those lenses. Even if the observation was difficult with the 20-foot and the 34-foot lenses, he had better optics in the following years, with which he was certainly able to resolve it.

\acknowledgments     
 
The authors deeply thank LESIA and Paris Observatory to allow them to do this study. They also want to thank Onera/DOTA/HRA, especially Bruno Fleury and Jean-Fran\c{c}ois Sauvage, to help us use their Zygo. Finally, they thank the Institut d'Optique, especially Lionel Jacubowiez and Thierry Avignon to give us access to their equipment.

Figures~\ref{fig:cassini}, \ref{fig:obs-paris} and \ref{fig:objectifs} are copyright of the Library of Paris Observatory, thanks are given for permission to include in this paper.


\bibliography{report}   
\bibliographystyle{spiebib}   

\end{document}